\documentstyle[prd,aps]{revtex}
\begin{document}
\draft

\def\beq{\begin{equation}}
\def\eeq{\end{equation}}
\def\bea{\begin{eqnarray}}
\def\eea{\end{eqnarray}}
\def\simlt{\stackrel{<}{{}_\sim}}
\def\simgt{\stackrel{>}{{}_\sim}}
\input epsf
\twocolumn[\hsize\textwidth\columnwidth\hsize\csname
@twocolumnfalse\endcsname

\title{Comment on ``Self-interacting Warm Dark Matter''}
\author{Fernando Atrio Barandela$^{(1)}$, Sacha  Davidson$^{(2)} $}

\address{$^{(1)}${\it Salamanca }}
\address{$^{(2)}${\it Theoretical Physics,  Oxford
University, 1 Keble Road, Oxford, OX1 3NP, UK }}

\date{November, 2000}
\maketitle
\begin{abstract}
Hannestad and Scherrer analysed self-interacting Warm and Hot
Dark Matter and found power spectra in disagreement with
previous papers. We argue that they 
include self-interactions in a weak-coupling approximation,
as for photons after recombination.
The ``tightly coupled'' approximation
used for pre-recombination photons, and previous discussions
of self-interacting HMD, would have been more appropriate.
Their approximation  generates a Boltzmann hierarchy
leading to a stiff system of equations. Furthermore
contamination by gauge modes could invalidate their results. 
\end{abstract}
\pacs{PACS:95.35.+d, 14.80.-j, 98.65.Dx  }
\vskip2pc]

\def\simlt{\stackrel{<}{{}_\sim}}
\def\simgt{\stackrel{>}{{}_\sim}}

Neutrinos with strong self-interactions
were originally suggested as a
dark matter candidate\footnote{a light
bosonic particle could behave very differently \cite{Toni}.}
by Raffelt and Silk
\cite{RS}, who estimated that density perturbations in such a fluid
would damp by diffusion (``Silk damping''), rather
than by free-streaming. In
a subsequent analysis---partially numerical, partially
analytic---Atrio-Barandela and Davidson (ABD)
\cite{ABD} computed a power spectrum in agreement
with this expectation. More recently,
power spectra were computed by
Hannestad and Scherrer \cite{H+S} for
self-interacting hot and warm dark matter, using
CMBFAST \cite{CMBFAST}. They find that the
shape of the power spectrum is essentially independent of
the strength of the dark matter self-interactions,
but depends on the mass of the dark matter particle, 
suggesting that the damping of density fluctuations
is the due to free-streaming rather than  diffusion.  
Their power spectra do not show the damping
due to diffusion and the
short distance oscillations
that one would expect for an interacting 
dark matter fluid \cite{RS,ABD}. We believe this
is due to an inappropriate analytic 
approximation, and possibly to contamination
by gauge modes. 

HS study the evolution of perturbations
$\Psi(x^i,q_i,\tau)$ in the dark matter
phase space distribution $f$:
\beq
f(x^i,q_i,\tau)  = f_0 (1 + \Psi(x^i,q_i,\tau))
\label{aHS}
\eeq
$f_0$ is the homogeneous and isotropic 
equilibrium phase
space distribution,
and $x^i,\tau$ are comoving coordinates
(see \cite{MB} for notation). In the absence of interactions,
$f$ satisfies the Boltzman equation for free
particles
\beq
{\cal L}[f] = \frac{ \partial f}{\partial \tau} +
 \frac{ \partial x^i}{\partial \tau}  \frac{ \partial f}{\partial x^i} +
\frac{ \partial q_j}{\partial \tau}  \frac{ \partial f}{\partial q_j} = 0 
\eeq
The CMBFAST  Boltzman code solves this
equation for $\Psi$ to evolve the HDM density
perturbations. 

The Boltzman equation
for an interacting species is
\beq
{\cal L}[f] = {\cal C}[f]
\eeq
where the collision term
${\cal C}[f]$ is difficult to
solve.
Schematically:
\beq
 {\cal C}[f_a]= \frac{1}{2E_a} \int d \Pi_b d \Pi_c d \Pi_d 
|{\cal M} |^2 \hat{\delta} (f_a f_b - f_c f_d)
\eeq 
where $d \Pi_b = d^3p_b/[(2 \pi)^3 2 E_b]$,  ${\cal M} $
is the matrix element for the process $a+b \leftrightarrow c+d$,
$\hat{\delta}$ is the 4-momentum conservation delta-function,
and we have assumed low occupation numbers.
An {\it equilibrium} distribution is defined
to satisfy  $ {\cal C}[f]= 0$, so 
${\cal L}[f_0] =0 =  {\cal C}[f_0]$, and
\beq
{\cal L}[f_0 \Psi] = {\cal C}[f_0 \Psi]
\label{star}
\eeq
as noted by HS.  We first discuss an 
approximate solution 
suitable for strongly interacting
particles, and then compare to  the 
method of HS.

The timescale associated with
the LHS of equation (\ref{star}),
for horizon-sized modes, is the
age of the Universe $\tau_U$. The timescale
associated with the RHS
is the interaction timescale
$\tau_\nu$ and $\tau_\nu \ll \tau_U$ for
strong self-interactions. In this
case,  (\ref{star}) can be solved in
the ``tightly-coupled'' approximation used for
pre-recombination photons: perturbations will
be very close to locally in equilibrium, because
the interactions are fast. So
\beq
f(x^i,q_i,\tau) = g^{(0)}(x^i,\tau) + g^{(1)}(x^i,q_i,\tau)
\label{aABD}
\eeq
where
\beq
 g^{(0)}(x^i,\tau) = \exp\{(E - \mu(x^i,\tau))/T(x^i,\tau) \} ~~,
\eeq
is the locally in equilibrium  distribution, so
$g^{(1)}$ is  the perturbation
away from {\it local} equilibrium.
By definition $ {\cal C}[g^{(0)}]= 0$,
so the Boltzman equation is
\beq
{\cal L}[f_0 \Psi] = {\cal C}[g^{(1)}] \simeq \Gamma g^{(1)} 
~~(\Gamma = \tau_\nu^{-1})
\label{twostar}
\eeq
This implies $g^{(1)} \sim (\tau_\nu/\tau_U) f_0 \Psi$.
The equality in eqn (\ref{twostar}) is the analytic
approximation required
to generate a hierarchy of Boltzmann equations similar
to the one solved by HS. 
From the approximation in eqn  (\ref{twostar}),
 we showed 
that density perturbations in interacting
HDM models would 
be damped by diffusion \cite{ABD},
as anticipated in \cite{RS}.
(This is also the result  obtained  by HS in
their Appendix 2.)

HS used a different approach from us; they take 
\beq
{\cal C}[ \Psi] \simeq \Gamma  \Psi
\label{HS}
\eeq
and generate a Boltzmann hierarchy. The
resulting equations are, however,  very stiff. 
Their approximation (equation \ref{HS})
diverges in the limit of very large scattering cross
sections and will not model the fact
that fast interactions can bring $ \Psi$ into a
local equilibrium form almost satisfying ${\cal C}[ \Psi] = 0$.
Therefore, we believe
that equation (\ref{HS}) does not correctly model the HDM
self-interactions, and this could explain
why HS's power spectra look like
non-interacting Hot Dark Matter.
Equation (\ref{HS}) would be valid in  the weak coupling
approximation used for photons after recombination,
as opposed to the pre-recombination
tight coupling approximation leading to
equation (\ref{twostar}). 

Note that both ABD and HS work
in synchronous gauge, where
density perturbations in the
CDM regime are defined in the matter rest
frame. However, the
divergence of the fluid velocity $\theta$ is non-zero 
for relativistic neutrinos, so
one must verify that 
$\theta_\nu =0$ after the $\nu$ become
non-relativistic, to ensure that
unphysical gauge modes do not contaminate the solution.
This could be problematic in the full numerical approach
of HS. 

\begin{figure}[htb]
\begin{center}
\epsfxsize=8cm \epsfysize=5cm \epsfbox{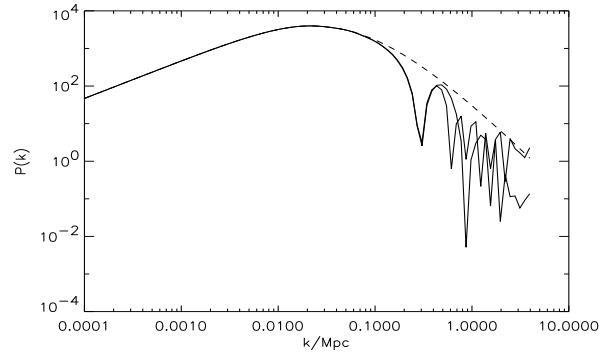}
\caption{ Power spectrum for two different
``sticky neutrino'' models. The solid lines
correspond (in decreasing amplitude)
to interactions 
with co-moving mean free paths at
$T_{\gamma} = 10$ eV of .01 Mpc and  .1 Mpc. These
give rise to estimated  damping scales of 7 Mpc$^{-1}$
and  3 Mpc$^{-1}$, respectively. The dashed line
correspond to standard CDM and is plotted for
comparison. The y-axis scale is arbitrary.
We took $H_o = 50$Km s$^{-1}$Mpc$^{-1}$.}
\label{fig}
\end{center}
\end{figure}

As mentioned, HS discuss the relation of
their paper to ABD in an Appendix, where 
they reproduce the analytic estimates \cite{RS,ABD}
showing that the perturbations should
oscillate within the horizon and
be damped on the diffusion scale. They
do not comment on the discrepancy
between the estimates and their plots. 
They suggest that the difference between the
ABD $P(k)$ plot  and theirs is due to
an extra $H \theta$ term appearing in our equations,
where $\theta$ is the divergence of the fluid velocity.
This term is due to a difference
in conventions: as stated in our paper,
we use the conventions of \cite{PC}  for
our analytic derivations, so 
$\theta = i k \cdot v/a$, and our
equations are correct (see
eqn 85.8 in \cite{PC}). HS use the conventions
of \cite{MB}, according to which
$\theta = i k \cdot v$, and the offending
 $H \theta$ term does not appear. 
In MB notation, where $\delta$
represents the density perturbation,
 equations ABD [40] and [41] are:
\bea
\dot\delta &=& -(1+w)(\theta+{1\over 2}\dot h)-3{\dot a\over a}({\dot P\over \dot \rho}-w)\delta\\
\dot\theta &=& -{\dot a\over a}(1-3w)\theta -{\dot w\over 1+w}\theta + k^2
{\dot P/\dot\rho\over 1+w}\delta -k^2\sigma(\theta-3\dot\eta)
\eea
where $w=P/\rho$, P =  pressure, $\rho = $  density, 
$a$ is the scale factor, $\sigma$ the viscosity,
$k$ the wavelength  in conformal units, $h$ and $\eta$ are the
metric perturbations, and derivatives are  with respect
to  conformal time. In the  relativistic
regime $w = 1/3, \dot w= 0$ and neutrinos behave like
a damped harmonic oscillator, while in the non-relativistic
regime density perturbations grow like Cold Dark Matter. 
 However, HS
are correct to doubt our plots; our treatment of the
transition from relativistic neutrino fluid  to
non-relativistic fluid was simplistic, 
and we did not correctly take into account the effect of
viscosity on small scales (a factor 3 was missing
in our programme).
During the transition period from radiation
to matter dominated regimes we assumed that $w=P/\rho$,
was linear with conformal time.
Other relationships gave very similar
power spectra.  We  have corrected
the viscosity term, and plot the power spectrum 
 in figure \ref{fig}.
This correction has
a negligeable effect: the small scale oscillations
are slightly reduced with respect to  \cite{ABD}, and our results agree
with expectations. 

\acknowledgements
We would like to thank Steen Hannestad
for discussions, and Joe Silk for useful  comments.

\end{document}